\PassOptionsToPackage{pdfpagelabels=false,hypertexnames=false}{hyperref}
\documentclass{IOS-Book-Article}
\usepackage[T1]{fontenc}
\usepackage{mathptmx}
\usepackage{graphicx,booktabs,microtype,amsmath,url,float,placeins}
\hypersetup{hidelinks}
\usepackage{flafter}
\usepackage{siunitx}
\DeclareSIUnit\pixel{px}
\usepackage[nameinlink,noabbrev]{cleveref}
\newcommand{\diceci}[2]{#1\,{\fontsize{7}{8}\selectfont$\pm \num{#2}$}}
\newcommand{\papertitle}{Does Vendor Image Processing Matter? Learning Echocardiography Display Appearance from Native Beamspace Data}
\newcommand{\firstauthorgiven}{Elias}
\newcommand{\firstauthorsurname}{Stenhede}
\newcommand{\secondauthorgiven}{Arian}
\newcommand{\secondauthorsurname}{Ranjbar}

\newcommand{\paperaffiliations}{Medical Technology \& E-Health, Akershus University Hospital, Norway}
\newcommand{\nTrain}{5943}
\newcommand{\nVal}{1176}
\newcommand{\nTest}{1287}

\newcommand{\maeInput}{33.85}

\newcommand{\maeLut}{5.42}

\newcommand{\gainMedOverLut}{43}

\newcommand{\paramsMed}{117k}

\newcommand{\cpuFps}{22}

\newcommand{\domainDelta}{9.3}
\newcommand{\domainDeltaCI}{8.8, 9.6}
\newcommand{\domainWins}{99}
\newcommand{\nSeeds}{3}

\newcommand{\gpuFps}{665}

\newcommand{\curveN}{50}
\newcommand{\oracleMae}{5.0}
\newcommand{\globalMaeFifty}{5.5}
\newcommand{\cnnMaeFifty}{3.3}

\newcommand{\taskThreeClips}{1125}
\newcommand{\taskThreeExams}{324}
\newcommand{\taskThreePatients}{318}

\newcommand{\matchedDiceDifferenceBound}{0.3}

\newcommand{\echoUnfilteredTrainCavityDrop}{4.3}

\newcommand{\echoUnfilteredTrainMyocardiumDrop}{11.4}

\newcommand{\echoFilteredTrainCavityDrop}{1.0}

\newcommand{\echoFilteredTrainMyocardiumDrop}{2.0}

\newcommand{\mae}{MAE}

\newenvironment{paperfigure}{\begin{figure}[H]}{\end{figure}}
\newenvironment{appearancetable}{\begin{table*}[!t]}{\end{table*}}

\newenvironment{camusfigure}{\begin{figure*}[!t]}{\end{figure*}}

\newenvironment{segmentationtable}{\begin{table*}[!t]}{\end{table*}}

\newenvironment{galleryfigure}{\begin{figure*}[!t]}{\end{figure*}}

\newcommand{\resultsbarrier}{\FloatBarrier}

\newenvironment{appearanceresults}{}{}

\renewenvironment{paperfigure}{\begin{figure}[!htbp]}{\end{figure}}
\renewenvironment{appearancetable}{\begin{table}[H]}{\end{table}}

\renewenvironment{camusfigure}{\begin{figure}[!htbp]}{\end{figure}}

\renewenvironment{segmentationtable}{\begin{table}[H]}{\end{table}}

\renewenvironment{galleryfigure}{\begin{figure}[!htbp]}{\end{figure}}

\begin{document}
\begin{frontmatter}
\title{\papertitle}
\author{\fnms{\firstauthorgiven} \snm{\firstauthorsurname}}
and
\author{\fnms{\secondauthorgiven} \snm{\secondauthorsurname}}
\address{\paperaffiliations}
\begin{abstract}
Most echocardiography datasets contain vendor-processed images, but processing is often undocumented and may affect model performance and robustness. To address this, we train a small convolutional neural network to reproduce General Electric display appearance from native beamspace brightness-mode ultrasound, using \num{8406} recordings from \num{605} unique patients. The beamspace CNN reduces test mean absolute error by \qty{\gainMedOverLut}{\percent} over the optimal pointwise brightness map, outperforms an equivalent Cartesian CNN and runs in real time on a CPU. Across the EchoXFlow and CAMUS datasets, unfiltered training followed by filtered evaluation yields the lowest mean Dice for every class, suggesting filtering may be useful as augmentation when evaluation data are filtered or their processing is unknown. Code and pretrained weights: \url{https://github.com/Ahus-AIM/EchoAppearance}.

\end{abstract}
\end{frontmatter}

\begin{paperfigure}\centering
\includegraphics[width=\linewidth]{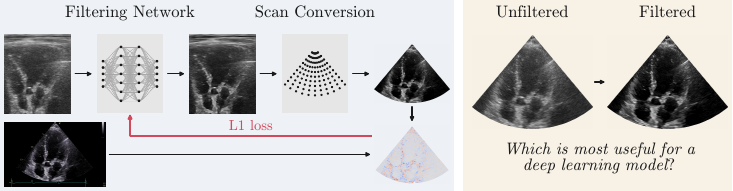}
\caption{Left: appearance learning against the vendor display. Right: the same frame without and with filtering.}\label{fig:pipeline}
\end{paperfigure}

\section{Introduction}
Images shown on a scanner's screen are processed versions of the native scan lines, and different vendors and presets give different looks. Echocardiography datasets are often shared after vendor-specific image processing, with limited documentation of the operations applied. As these datasets are combined to train deep learning models, such processing may influence downstream performance and robustness across imaging environments. We study distillation of display appearance: native beamspace brightness-mode (B-mode) data in, the display frame of a General Electric (GE) scanner out. We then assess how domain shifts caused by display processing affect segmentation performance across two datasets.

\section{Methods}
This section describes the three main parts of the article: frame registration, appearance filtering and downstream segmentation evaluation.

\subsection{Registration}
We use recordings from the EchoXFlow dataset~\cite{echoxflow}, acquired during routine clinical examinations at Akershus University Hospital, Norway, with a GE Vivid E95 Ultra Edition and a 4Vc 4D volume phased-array probe. Each recording is exported as beamspace B-mode (polar samples with the sector geometry) and the GE display frame, available as a JPEG of the first frame. Pixel-wise learning needs sub-pixel agreement, so every recording is registered to its display frame: we fit the sector and measure a patch-shift table using locally upsampled Fourier correlation~\cite{guizar2008registration}. A recording is kept when its patch shifts stay below \qty{0.25}{\pixel}, its fitted sector width is within \qty{0.2}{\percent} of the geometry. Recordings with colour flow or a doubtful frame match are excluded. Of \num{13389} eligible B-mode recordings, \num{8406} pass. Splits are by patient: \num{\nTrain} / \num{\nVal} / \num{\nTest} recordings. We scored the test split once, after freezing all models and hyper-parameters on validation.

\subsection{Filtering Network}
A residual CNN~\cite{he2016residual} (32 channels, 6 blocks of \numproduct{3x3} convolutions with Gaussian error linear unit (GELU) activations~\cite{hendrycks2016gelu}) runs on the native beam frame. A fixed bilinear resampling maps its output onto the display grid, where a small residual head (16 channels, one block) follows; the model has \paramsMed{} parameters. Initialised as the identity, it starts out as plain scan conversion (\Cref{fig:pipeline}). The loss is L1 on the display grid over supported pixels, so targets keep their original resolution. We train on \qty{128}{\pixel} patches with AdamMuon~\cite{jordan2024muon,loshchilov2019adamw}, halve the learning rate at plateaus and keep the best validation checkpoint.

The optimal pointwise map assigns each input grey level the median target value over the training recordings (monotone, 256 levels). It is the lowest mean absolute error (\mae{}) any brightness curve can reach (\Cref{fig:curve}).

The Cartesian twin has the same layers and parameters but applies them after scan conversion; only the order of resampling and convolution differs. We train three seeds of each. Metrics are masked \mae{}, peak signal-to-noise ratio (PSNR) and structural similarity (SSIM~\cite{wang2004ssim}); intervals use \num{2000} patient-bootstrap draws. With warmed batches of \num{32} frames (bfloat16), the \paramsMed{}-parameter beamspace model processes \numproduct{726x274} frames at \num{\cpuFps} frames per second on an AMD Ryzen Threadripper PRO 9965WX CPU (16 threads) and \num{\gpuFps} frames per second on an NVIDIA RTX 5090 graphics processing unit (GPU).

\subsection{Segmentation Network}
We train separate MViT-v2-S~\cite{li2022mvitv2} segmentation models on unfiltered and filtered Cartesian \numproduct{256x256} inputs, evaluating each checkpoint on both domains. The frozen appearance CNN operates in beamspace before scan conversion. Both conditions use identical clips, geometry, normalisation and training settings. EchoXFlow uses five patient-disjoint outer folds and three seeds, covering \num{\taskThreeClips} clips from \num{\taskThreeExams} examinations of \num{\taskThreePatients} patients. CAMUS~\cite{leclerc2019camus} uses its official 400/50/50 patient train/validation/test split and three seeds per condition. Training stops with validation patience 10 after at least 25 epochs; the best checkpoint is kept.

We report spatiotemporal Dice~\cite{dice1945} for the left ventricular (LV) cavity and myocardium, plus the left atrium on CAMUS. Scores are averaged within examinations and over seeds, then equally over EchoXFlow folds.

\suppressfloats[t]
\begin{appearanceresults}
\section{Results}
Scan conversion alone has a test \mae{} of \num{\maeInput}; the optimal pointwise map reduces it to \num{\maeLut} (\Cref{tab:main}), so most of the difference is a brightness mapping. The beamspace CNN is \qty{\gainMedOverLut}{\percent} lower than this map. The pointwise map darkens the input and keeps its spatial structure, while the beamspace CNN also changes the local texture (\Cref{fig:gallery}, insets). Fitting the pointwise map to each recording separately changes the curve moderately, but still reaches an \mae{} of \num{\oracleMae} on these \num{\curveN} validation recordings, against \num{\globalMaeFifty} for the global map and \num{\cnnMaeFifty} for the beamspace CNN.

\begin{appearancetable}\centering\footnotesize\setlength{\tabcolsep}{3pt}
\caption{Appearance fidelity on the held-out test set: recording means [\qty{95}{\percent} patient-bootstrap interval].}\label{tab:main}
\begin{tabular*}{\linewidth}{@{\extracolsep{\fill}}l*{3}{r@{\extracolsep{0pt}.}l@{\;}r@{,\,}l@{\extracolsep{\fill}}}@{}}\toprule
Method & \multicolumn{4}{c}{\mae{} $\downarrow$} & \multicolumn{4}{c}{PSNR (\unit{\decibel}) $\uparrow$} & \multicolumn{4}{c}{SSIM $\uparrow$} \\
\midrule
Scan conversion (input) & 33 & 8 & [33.6 & 34.1] & 17 & 1 & [17.1 & 17.2] & 0 & 512 & [0.498 & 0.525] \\
Optimal pointwise map & 5 & 4 & [5.2 & 5.6] & 29 & 5 & [29.3 & 29.8] & 0 & 892 & [0.887 & 0.896] \\
Cartesian CNN & 3 & 4 & [3.2 & 3.6] & 34 & 2 & [33.9 & 34.5] & 0 & 939 & [0.934 & 0.943] \\
\textbf{Beamspace CNN} & \textbf{3} & \textbf{1} & \textbf{[2.9} & \textbf{3.3]} & \textbf{35} & \textbf{2} & \textbf{[34.8} & \textbf{35.4]} & \textbf{0} & \textbf{945} & \textbf{[0.939} & \textbf{0.949]} \\
\bottomrule\end{tabular*}

\end{appearancetable}

\begin{galleryfigure}\centering
\includegraphics[width=\textwidth]{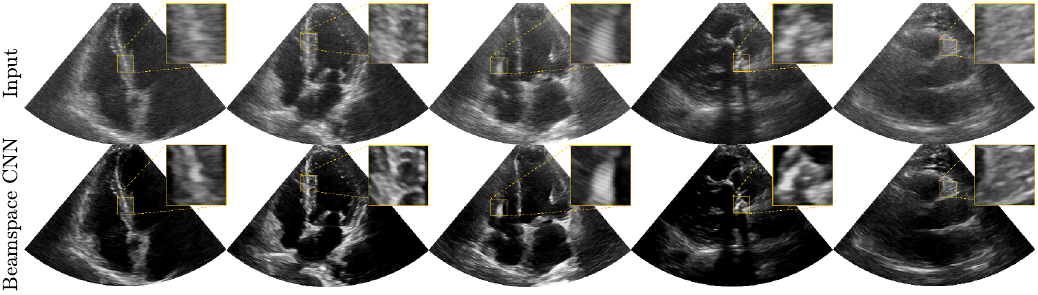}
\caption{Five test recordings from different patients, selected at fixed CNN-error quantiles. Rows: input and beamspace CNN.}\label{fig:gallery}
\end{galleryfigure}
\end{appearanceresults}

\begin{paperfigure}\centering
\includegraphics{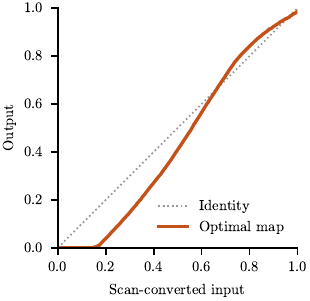}
\caption{The optimal pointwise map (median target per input level), fitted on the training recordings.}\label{fig:curve}
\end{paperfigure}

With layers, parameters and optimiser held equal, the beamspace model has \qty{\domainDelta}{\percent} lower \mae{} than its Cartesian twin, averaged over \num{\nSeeds} seeds each (\qty{95}{\percent} interval of the paired per-recording difference $[\domainDeltaCI]$\,\unit{\percent}; better on \qty{\domainWins}{\percent} of recordings). Inference uses the native beam data and the geometry.

With matched training and evaluation, mean Dice differs by at most \num{\matchedDiceDifferenceBound} percentage points (\Cref{tab:task3}). Filtering only at evaluation reduces EchoXFlow cavity and myocardium Dice by \num{\echoUnfilteredTrainCavityDrop} and \num{\echoUnfilteredTrainMyocardiumDrop} points, respectively. Removing it only at evaluation reduces Dice by \num{\echoFilteredTrainCavityDrop} and \num{\echoFilteredTrainMyocardiumDrop} points. CAMUS results are directionally similar but smaller, and not all changes are statistically significant.

\begin{segmentationtable}\centering\footnotesize\setlength{\tabcolsep}{3pt}
\caption{Dice (\unit{\percent}) $\uparrow$: mean $\pm$ conservative \qty{95}{\percent} CI.}\label{tab:task3}
\begin{tabular*}{\linewidth}{@{\extracolsep{\fill}}lccccc@{}}
\toprule
& \multicolumn{2}{c}{EchoXFlow} & \multicolumn{3}{c}{CAMUS} \\
\cmidrule(lr){2-3}\cmidrule(lr){4-6}
Filtering applied & LV cavity & Myocardium & LV cavity & Myocardium & Left atrium \\
\midrule
None & \diceci{\textbf{93.5}}{0.3} & \diceci{\textbf{76.0}}{0.8} & \diceci{93.8}{0.6} & \diceci{88.1}{1.2} & \diceci{92.2}{0.8} \\
Training and evaluation & \diceci{93.4}{0.3} & \diceci{75.7}{0.8} & \diceci{\textbf{93.9}}{0.6} & \diceci{\textbf{88.3}}{1.1} & \diceci{\textbf{92.4}}{0.9} \\
\addlinespace[3pt]
Evaluation only & \diceci{89.2}{0.6} & \diceci{64.5}{1.1} & \diceci{93.2}{1.0} & \diceci{87.3}{1.5} & \diceci{91.3}{1.0} \\
Training only & \diceci{92.4}{0.4} & \diceci{73.7}{0.8} & \diceci{93.6}{0.6} & \diceci{87.9}{1.1} & \diceci{92.4}{0.9} \\
\bottomrule
\end{tabular*}

\end{segmentationtable}

In the five CAMUS examples, filtering darkens the chambers and changes tissue texture (\Cref{fig:camus}).

\begin{camusfigure}\centering
\includegraphics[width=\linewidth]{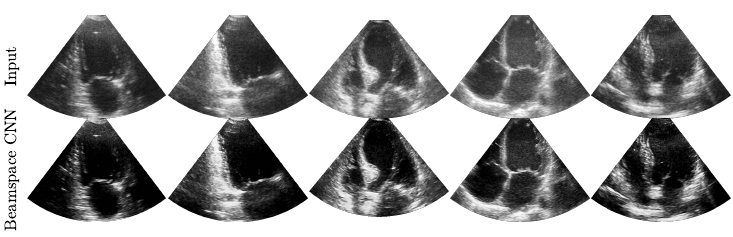}
\caption{External application to CAMUS: five patients. Top: input; bottom: beamspace CNN.}\label{fig:camus}
\end{camusfigure}

\resultsbarrier
\section{Limitations and Future Directions}
Targets are lossy JPEGs of one frame per recording from one scanner (GE Vivid E95 Ultra Edition) and probe (4Vc). The model reproduces observed appearance.

Future work could evaluate appearance filtering in other echocardiographic tasks, including image-based probe guidance~\cite{stenhede2026echoposed} and the training of foundation models~\cite{munim2026echojepa}.

\section{Conclusions}
A small beamspace CNN approximates GE display appearance efficiently and improves fidelity over pointwise mapping and a matched Cartesian CNN. Segmentation degrades when preprocessing changes between training and evaluation, while matched training yields small differences. More broadly, this work highlights the need for considering postprocessing settings before feeding data into deep learning systems, and provides a CNN for approximating one kind of vendor-specific postprocessing.

\FloatBarrier

\end{document}